\documentclass[11pt]{article}
\pdfoutput=1
\usepackage[utf8]{inputenc}
\usepackage[left=20mm,right=20mm,top=25mm,bottom=25mm]{geometry}

\usepackage[affil-it]{authblk}
\usepackage{graphicx}
\usepackage{amsmath}
\usepackage{amsfonts}
\usepackage{amssymb}
\usepackage{subcaption}

\usepackage{pgfplots}
\pgfplotsset{compat=newest} 
\pgfplotsset{plot coordinates/math parser=false} 

\pgfplotsset{every axis/.append style={
   legend style={nodes={font=\tiny}} 
  		},
		every axis/.append style = {nodes={font=\tiny}},
		every node/.append style={font=\tiny, anchor=north west},
		every node/.style={scale=0.7};
		nodes={font= \tiny}
   }
\usepackage{booktabs} 
\newcommand{\ra}[1]{\renewcommand{\arraystretch}{#1}}

\usepackage[german=quotes]{csquotes}
\usepackage{siunitx}

\usepackage[
colorlinks=true,
linkcolor=tud2ad,
anchorcolor=black,
citecolor=tud2ad,
filecolor=magenta,
menucolor=red,
urlcolor=tud2ad,
plainpages=false,
pdfpagelabels,
hypertexnames=false,
linktocpage
]{hyperref}

\definecolor{fnbBlau}{HTML}{009CDA}
\definecolor{tud1a}{HTML}{5D85C3}
\definecolor{tud2a}{HTML}{009CDA}
\definecolor{tud2ad}{rgb}{0.00000,0.44700,0.74100}%
\definecolor{tud3a}{HTML}{50B695}
\definecolor{tud4a}{HTML}{AFCC50}
\definecolor{tud5a}{HTML}{DDDF48}
\definecolor{tud6a}{HTML}{FFE05C}
\definecolor{tud6b}{HTML}{FDCA00}
\definecolor{tud8a}{HTML}{EE7A34}
\definecolor{tud8b}{HTML}{EC6500}
\definecolor{tud9a}{HTML}{E9503E}
\definecolor{tud11b}{HTML}{721085}
\definecolor{tud1b}{HTML}{005AA9} 
\definecolor{tud3b}{HTML}{009D81}
\definecolor{tud4b}{HTML}{99C000}
\definecolor{tud4c}{HTML}{7FAB16}
\definecolor{tud4d}{HTML}{6A8B22}
\definecolor{tud9b}{HTML}{E6001A}
\definecolor{tud9c}{HTML}{B90F22} 
\definecolor{sourcecode}{HTML}{E9E9E9}

\newcommand{\Arch}{\operatorname{\mathit{A\kern-.06em r}}} 
\newcommand{\Biot}{\operatorname{\mathit{B\kern-.06em i}}} 
\newcommand{\Cauc}{\operatorname{\mathit{C\kern-.07em a}}} 
\newcommand{\Damk}{\operatorname{\mathit{D\kern-.06em a}}} 
\newcommand{\Eule}{\operatorname{\mathit{E\kern-.03em u}}} 
\newcommand{\Four}{\operatorname{\mathit{F\kern-.10em o}}} 
\newcommand{\Frou}{\operatorname{\mathit{F\kern-.07em r}}} 
\newcommand{\Gras}{\operatorname{\mathit{G\kern-.05em r}}} 
\newcommand{\Karl}{\operatorname{\mathit{K\kern-.11em a}}} 
\newcommand{\Knud}{\operatorname{\mathit{K\kern-.11em n}}} 
\newcommand{\Lewi}{\operatorname{\mathit{L\kern-.05em e}}} 
\newcommand{\Mach}{\operatorname{\mathit{M\kern-.10em a}}} 
\newcommand{\Nuss}{\operatorname{\mathit{N\kern-.09em u}}} 
\newcommand{\Pecl}{\operatorname{\mathit{P\kern-.08em e}}} 
\newcommand{\Pran}{\operatorname{\mathit{P\kern-.03em r}}} 
\newcommand{\Rayl}{\operatorname{\mathit{R\kern-.04em a}}} 
\newcommand{\Reyn}{\operatorname{\mathit{R\kern-.04em e}}} 
\newcommand{\Rich}{\operatorname{\mathit{R\kern-.06em i}}} 
\newcommand{\Schm}{\operatorname{\mathit{S\kern-.07em c}}} 
\newcommand{\Sher}{\operatorname{\mathit{S\kern-.07em h}}} 
\newcommand{\Stro}{\operatorname{\mathit{S\kern-.07em r}}} 
\newcommand{\Webe}{\operatorname{\mathit{W\kern-.14em e}}} 

\title{Autonomous Cooking with Digital Twin Methodology}
\date{March 11, 2021}
\author[1,*]{Maximilian~Kannapinn}
\author[1]{Michael~Schäfer}
\affil[1]{\footnotesize Institute for Numerical Methods in Mechanical Engineering \& Graduate School of Computational Engineering, \protect \\
Department of Mechanical Engineering \& Centre for Computational Engineering,\protect \\ Technical University of Darmstadt, Dolivostr.~15, 64293 Darmstadt, Germany}
\affil[*]{\footnotesize Corresponding author, Email: \href{mailto:research@maxkann.de}{research@maxkann.de}}

\begin{document}

\maketitle
 \par\noindent\rule{\textwidth}{0.4pt}           
\begin{abstract} \noindent
This work introduces the concept of an autonomous cooking process based on Digital Twin methodology. It proposes a hybrid approach of physics-based full order simulations followed by a data-driven system identification process with low errors. It makes faster-than-real-time simulations of Digital Twins feasible on a device level, without the need for cloud or high-performance computing. The concept is universally applicable to various physical processes.
\end{abstract}

\vspace*{0.5ex}
\par\noindent\rule{\textwidth}{0.4pt}\vspace*{2pt}
{\small
Accepted version of manuscript published in \emph{Proceedings of WCCM-ECCOMAS 2020}. \\
Date accepted: March 11, 2021. DOI: \href{https://doi.org/10.23967/wccm-eccomas.2020.074}{10.23967/wccm-eccomas.2020.074}. 
License: \href{https://creativecommons.org/licenses/by-nc-nd/4.0/legalcode}{CC BY-NC-ND 4.0}
}
\vspace*{-1.6mm}
\par\noindent\rule{\textwidth}{0.4pt}


\newlength\fheight
\newlength\fwidth
\newlength\figureheightbig
\newlength\figurewidthbig

\section{INTRODUCTION}

Autonomous processes are without question the next big disruptive technology trend. Ambitious self-driving car projects by major tech companies demonstrate the progress industry has made in the past decade. 
In contrast to these well-known endeavours, the present work sheds light on the yet unconsidered potential of autonomous cooking processes through Digital Twin (DT) technology. The automation of food processing does not only imply natural industrial benefits but, more importantly in modern times, environmental and health aspects on larger scales as well.
Intelligent cooking devices may be beneficial in the quest to transform our food system to help us evolve towards a more environmentally-friendly future. Following the EU Farm to Fork Strategy and Circular Economy Action Plan, we could reach the sustainability goals of the European Green Deal 2030~\cite{mot_greendeal2030}. It becomes clear that a change in our food system towards less wastage can contribute to our strive to keep global temperatures at safe levels. For instance, a recent Special Report on Climate Change and Land of the Intergovernmental Panel on Climate Change (IPCC) 
attributed eight to ten percent of the total anthropogenic greenhouse gas emissions to global food loss and wastage~\cite{mot_ipcc_climate2019_short}.
Besides the impact on Climate Change, it is imperative to reach the sustainable development goals of the United Nations, e.g. zero hunger~\cite{mot_unitednationsgoal}. 
Especially developing countries require safe and healthy meals in large amounts. Wasting of foods can be related here to the lack of proper processing and preservation techniques and facilities~\cite{mot_fao_food2011}.

In general, autonomous cooking processes can improve our overall food safety, as they may ensure a more reliable neutralisation of bacterial pathogens.
Langsrud~\cite{mot_langsrud_cooking2020} disclosed  concerning deficits in our approaches to judge the doneness of our food: 
\enquote{About one third of foodborne illness outbreaks in Europe are acquired in the home and eating undercooked poultry is among consumption practices associated with illness.}
Food processing of large quantities while preserving food quality is vital for community catering, e.g. crisis help, canteens, schools and universities. Although cooking needs experience, workforce and time, we face a shortage of skilled staff in recent years. Oberhuber~\cite{mot_zeit2018} concludes that chef and sous-chef were ranked fourth and sixth of the most unpopular jobs in Germany in 2018. Every third business in gastronomy has difficulties in finding employees. The number of apprentices dropped to the lowest level since 1976 and every second trainee quits the program~\cite{mot_zeit2018}.

The scenarios mentioned above emphasise some of the potentials of autonomous cooking processes. In summary, we strive for positive progress
in terms of safe and healthy food production. It is vital to meet ambitious sustainability goals, minimise the food value chain’s energy consumption and reduce food wastage during industrial processing. 
This work is applied research enabling fast transfer to an industrial level. 
In the following section, we formulate a hybrid physics-based data-driven Digital Twin framework for autonomous cooking processes. In section~3, a model of hygroscopic capillary-porous food is introduced. Our implementation is benchmarked with results of Ni~\cite{pm_ni_multiphase}. Section 4 focusses on data-based non-linear system identification of the model. We demonstrate how CPU costs are reduced significantly. It enables us to predict several possible future cooking control paths. In section 5, we summarise our work and give an outlook on the coupled simulation of food processing devices and cooking physics.

\section{A HYBRID PHYSICS-BASED DATA-DRIVEN DIGITAL TWIN}
\label{sec:hybridtwin}

The overall premise of making a process controllable, or even autonomous, is gathering real-time information on its current state. Sensory equipment cannot capture all relevant information easily or feasibly. Sometimes it is not possible at all. Considering autonomous cooking, we may measure the core temperature of our food with simple thermometers. However, cooking devices cannot measure the food’s sensory properties such as the Mailliard browning progress (influencing colour and flavour), texture (e.g. tenderness) or residual moisture content. 
Hence, it is an open research question on how to gather process-critical information efficiently and precisely to perform decision making during live operations. It is a substantial criterion of autonomy to obtain and evaluate this information with none or minimal user-interaction. 

\subsection{Definition of Digital Twins}

In the past years, the concept of a DT has been identified as a possible framework to enable unknown state information acquisition. Tao’s and Qi’s “Make more digital twins” article in Nature 2019~\cite{dt_tao_nature} puts the focus on DTs 
as a prosperous research topic at the latest.
After Grieves’ first mention of DTs, their definition has been interpreted, extended and modified by the research community. 
For clarity, we will give a summary of our definition hereafter.

\begin{figure}[htbp]
 \centering
\includegraphics[width=0.9\textwidth]{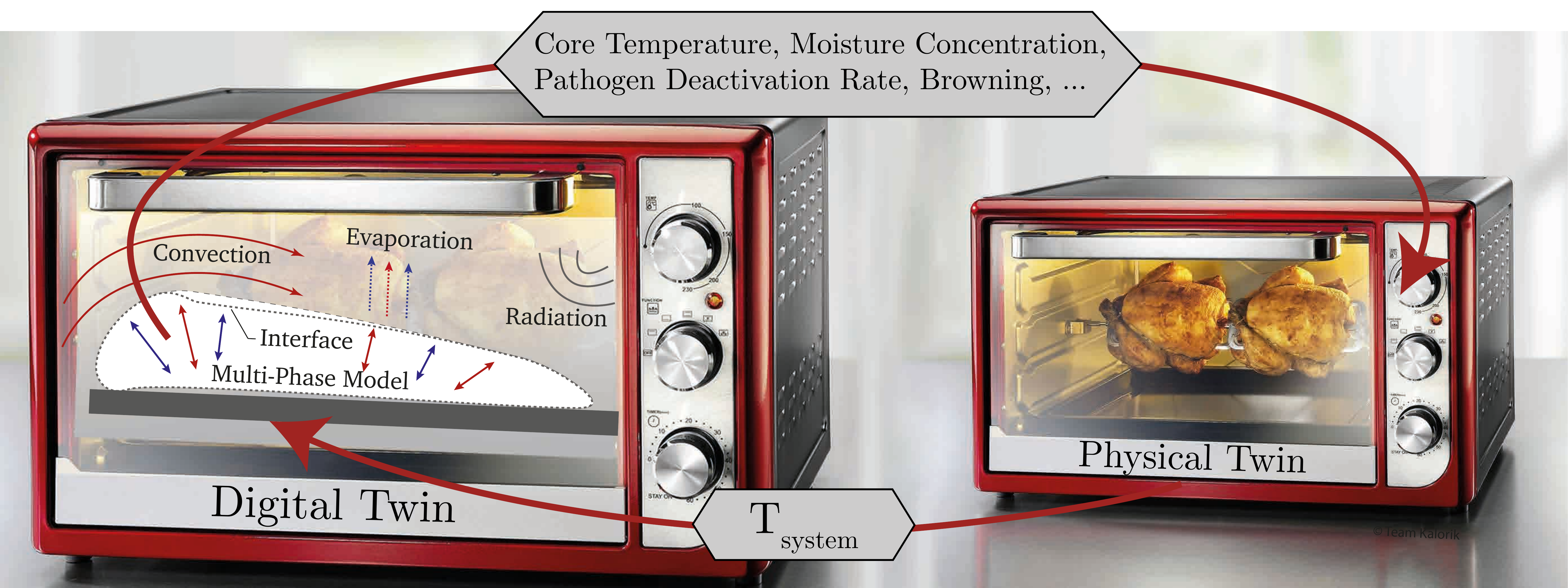}
 \caption{Digital Twins for autonomous cooking processes.}
 \label{fig:dt_microwave}
\end{figure}

The essential notion of a DT is to set up a virtual doppelganger simulation of the real-world process. 
Fig.~\ref{fig:dt_microwave} illustrates how simple measurements of the physical process, in our case temperatures of food or oven walls, are transferred to the DT. This data forms the boundary and initial conditions of the simulation model. 
Previously unknown state information, such as brownness or tenderness, is generated through simulation and is then handed back to the device’s control system.
The use of DTs during operation implies the need for rapid simulation results on the virtual side. We motivate the need for faster-than-real-time simulations in the following section.

\subsection{Need for faster-than-real-time simulation}
The novelty of the DT concept lies in bi-directional data exchange between the real-world object and its digital counterpart in real-time. Without this notion, every simulation model could be a DT. Unfortunately, the latter has been practised over the past years occasionally, leading to an ambiguous perception of the concept itself. This work stretches the necessity to outreach the minimum requirement of real-time simulations to make autonomous processes possible. 
Although real-time data provision permits closed-loop control algorithms, this is not sufficient for an autonomous process design. It is rather mandatory to predict multiple future scenarios to optimise future operational behaviour. Commonly known techniques are optimal control and model predictive control. An example of a global optimum criterion could be: \textit{Prepare medium-rare meat, tender, light browning and reach ready-to-eat temperatures within precisely 15 minutes}. 

Although we saw substantial improvements in hardware (e.g. FLOPS) and software (e.g. solver algorithms and parallelisation) for simulation, we are -- in general -- still not able to simulate large, coupled systems in real-time. It can be assumed that, from today’s point of view, new imperatives like high performance or cloud computing will not enable real-time or faster simulations for these models in the foreseeable future.

\subsection{Obtaining high accuracy with physics-based full order models}

Physically detailed models can generate high-quality training data. Covering a wide parameter range provides deep insights into the coupling of heating equipment and cooking physics. 
The promotion of faster-than-real-time simulation on the virtual side inherently requires a strategy to reduce the computational cost at some point. 
We advocate with our hybrid concept to not trade-in complexity for performance during physical modelling. We believe that the negative impact on precision is more prevalent here than at later stages, e.g. during data-driven system identification. The loss of accuracy is minimal there, as demonstrated in Sec.~\ref{sec:dynrom}.
Our offline approach does not need to make compromises in CPU capacity, model complexity or time constraints. Simulations can meet custom quality criteria or follow the state-of-the-art. 

\subsection{Digital Twin based autonomous cooking framework} 

\begin{figure}[hbtp]
 \centering
\includegraphics[width=\textwidth]{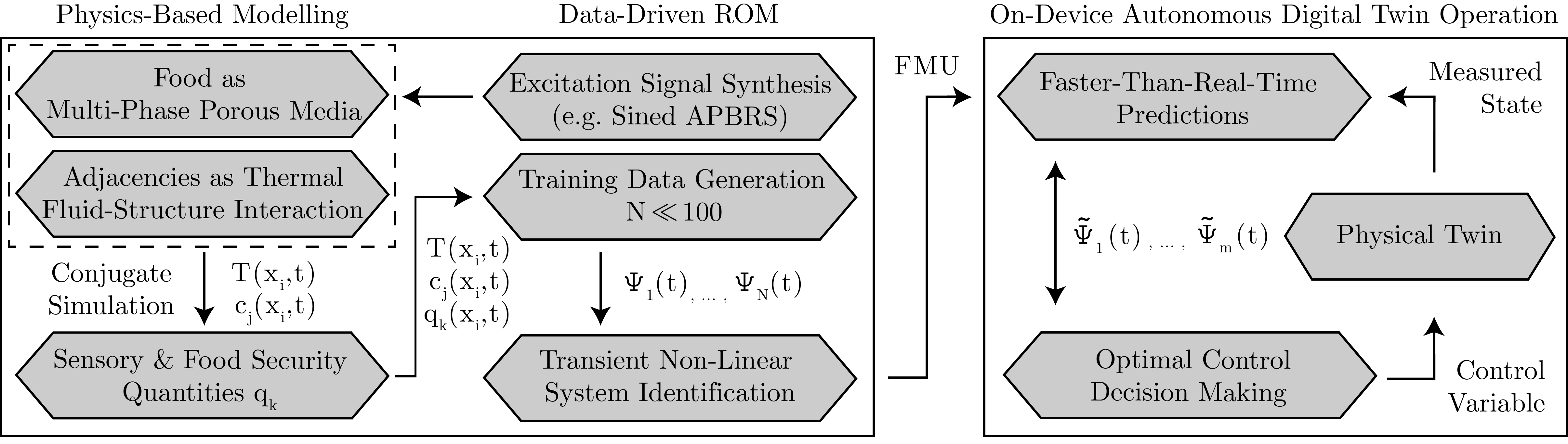}
 \caption{Concept of hybrid physics-based data-driven Digital Twin framework.}
 \label{fig:dt_famework21}
\end{figure}

Coming from the above-mentioned requirements, Fig.~\ref{fig:dt_famework21} shows our derived framework to design an autonomous cooking process with a DT methodology.

Detailed physics-based simulations form our basis. 
Models of the food and its physical and chemical processes consist of primary and secondary quantities. Primary data, dependent in space ($x_i$) and time ($t$), can be considered temperature $T\left(x_i,t\right)$ and moisture concentration $c_\text{w}\left(x_i,t\right)$. Secondary quantities $q_k$ like sensory attributes or pathogen deactivation rates can be modelled with Arrhenius-type ODEs~\cite{cm_rabeler_kin2018}.
The surrounding process conditions are characterised by fluid flow (for buoyancy and forced convection)~\cite{pm_datta_toward2016}, thermodynamics (conjugate heat transfer and radiation)~\cite{ir_almeida_measurement2006} and electromagnetics (e.g. microwave heating~\cite{pm_kumar_microwave}).
The adjacencies have to be coupled with the food cooking process to link the product and process mechanisms. 

A data-driven non-linear system identification process performs the reduction to a Reduced-Order Model (ROM), compare Sec.~\ref{sec:dynrom}. Training data $\Psi_i$ is obtained by evaluating the full-order model with a wide parameter range for the input (excitation) signal, e.g. the applied heat flux to the cooking process. 
Model exchange standard formats like Functional Mockup Unit (FMU) contain the identified ROM. It can be executed live to produce $m$ estimates of the state $\widetilde{\Psi}_i$ on the device with minimum system requirements, compare Tab.~\ref{tab:predictionperformance} in Sec.~\ref{sec:performanceanalysis}.

\section{FOOD AS POROUS MEDIA} \label{sec:porousmedia}

Food processing can be modelled as multi-phase heat and mass transfer in hygroscopic porous media. The following model is based on well-validated works of Datta~\cite{ pm_datta_porous2007a,pm_datta_toward2016}. 
It is not feasible to simulate gases and liquids within the porous domain’s exact representations. All quantities of interest are averaged on a sufficiently large Representative Elementary Volume (REV). 
In order to formulate relations that are analogous to well-established, non-hygroscopic porous media equations, a constant equivalent porosity
\begin{align}
\phi = \phi_\text{g} + \phi_\text{w} = \frac{V_\text{g}}{V_\text{tot}} + \frac{V_\text{w}}{V_\text{tot}}
\end{align}
is chosen as the sum of gas and water porosity. Porosity is defined as the ratio of a fluid's volume $V_\text{i}$ (i $=$ gas, water, vapor) to the total volume $V_\text{tot}$ of the REV. To account for porosity changes due to processing, equivalent variable water and gas saturations 
\begin{align}
S_\text{w} = \frac{V_\text{w}}{V_\text{w} + V_\text{g}} = \frac{V_\text{w}}{\phi \,V_\text{tot} } \quad \text{and} \quad
S_\text{g} = \frac{V_\text{g}}{V_\text{w} + V_\text{g}} = \frac{V_\text{g}}{\phi \,V_\text{tot} }
\end{align}
are defined. The saturations represent the relative volume fraction of the corresponding fluid in the porous cavity, resulting in $S_\text{w}+S_\text{g} = 1$.
The concentrations of water, gas (vapor + air) and vapor are defined as
\begin{align}
c_\text{w} = S_\text{w} \, \phi \,\rho_\text{w}\,, \qquad
c_\text{g} = S_\text{g} \, \phi \, \rho_\text{g} = S_\text{g} \, \phi \, \frac{p\, M_\text{g}}{R\,T} \,,\qquad
c_\text{v} = S_\text{g} \, \phi \, \rho_\text{v}  \,,
\end{align}
where the densities $\rho_i$ are determined employing Ideal Gas law, $M_i$ represent the molecular weights and $R$ is the universal gas constant. 

\subsection{Conservation laws}
The conservation of mass for gas, vapor and water is modelled by the following convection-diffusion equations:
\begin{align}
\frac{\partial c_\text{g}}{\partial t}+\nabla \cdot \left( -\rho_\text{g} \frac{k_\text{g}}{\mu_\text{g}} \nabla p \right)&=\dot{I}\,, \\
\frac{\partial c_\text{v}}{\partial t}+\nabla \cdot \left( -\rho_\text{v} \frac{k_\text{g}}{\mu_\text{g}} \nabla p -\phi \,S_\text{g}\, \rho_\text{g}\, D_\text{eff,g} \nabla \omega_\text{v} \right) &=\dot{I} \,,\\
\frac{\partial c_\text{w}}{\partial t}+\nabla \cdot\left( -\rho_\text{w} \frac{k_\text{w}}{\mu_\text{w}}\nabla p - D_{\text{w}, c_\text{w}} \nabla c_\text{w} \right)&=-\dot{I}\,.
\end{align}
Three major transport mechanisms can be identified: mass fluxes due to pressure gradients can be approximated (for $\Reyn < 1 \sim 10$) applying Darcy's law to obtain $\vec{j} = - \rho_i \frac{k_i}{\mu_i}\,\nabla p$, where $k_i$ is permeability, $\mu_i$ is the dynamic viscosity and $p$ is pressure. 
For vapor, we account for additional diffusive fluxes due to binary diffusion, where $\omega_\text{v}$ is the mass fraction of vapor and $D_\text{eff,g}$ represents the effective gas diffusion coefficient~\cite{pm_bird_transport}. Unsaturated capillary flow of water tends to move from locations with higher concentrations to ones of lower concentrations which is reflected in a variable capillary flow coefficient $D_{\text{w}, c_\text{w}}$. The reader is referred to works by Datta~\cite{pm_datta_porous2007a,pm_datta_toward2016} for further details and derivations.

Energy is conserved considering convective and diffusive mass fluxes
\begin{align}
\left(\rho c_\text{p} \right)_\text{eff} \frac{\partial T}{\partial t}+\sum_{i=\text{g,w}} \vec{j}_{i} \cdot \nabla\left(c_{\text{p},i} \,T\right)&=\nabla \cdot\left(k_\text{eff} \,\nabla T\right)-\lambda \dot{I} \,,
\end{align}
where $\lambda$ is latent heat of evaporation and effective transport coefficients are averaged via
\begin{align}
\left(\rho c_\text{p} \right)_\text{eff} &= \rho_\text{s}\,(1-\phi)\,c_\text{p,s} + \rho_\text{g}\,S_\text{g}\,\phi\,c_\text{p,g} + \rho_\text{w}\,S_\text{w}\,\phi\,c_\text{p,w} \,,\\
k_\text{eff} &= k_\text{s}\,(1-\phi) + k_\text{g}\,S_\text{g}\,\phi + k_\text{w}\,S_\text{w}\,\phi \,. \label{eq:keff}
\end{align}

The non-equilibrium formulation for distributed evaporation processes 
\begin{equation}
\dot{I} = K_\text{evap}\, ( \rho_\text{v,equ.} - \rho_\text{v})\,S_\text{g}\,\phi
\end{equation}
closes the set of equations. The evaporation constant $K_\text{evap}$ is the inverse of the time to reach evaporation equilibrium at molecular level and $\rho_\text{v,equ.} = a_\text{w}\,p_\text{sat} \,\frac{M_\text{v}}{R\,T}$ is the the equilibrium vapor density. It relates to the specific food material's water activity $a_\text{w}$ and vapor saturation pressure $p_\text{sat} (T)$.

\subsection{Test case}

Let us consider a slice of food with thickness of \SI{0.01}{m} on a convection oven's baking plate. Ignoring corner effects in horizontal directions, we can reduce the problem to a 1D formulation for simplicity. The only interface to its surroundings is the surface ($y=\SI{0}{m}$), where boundary conditions are prescribed:
\begin{align}
\left. p \right|_\text{surf} &= p_\text{amb}\,, \\ 
\left. \vec{j}_\text{v} \right|_\text{surf} &= h_\text{m} \, \phi \, S_\text{g} \left( \rho_\text{v} - \rho_\text{v,\text{oven}}\right) ,\\
\left. \vec{j}_\text{w} \right|_\text{surf} &= h_\text{m} \, \phi \, S_\text{w} \left( \rho_\text{v} - \rho_\text{v,\text{oven}}\right) ,\\ 
\left. \vec{q}\, \right|_\text{surf} &= h_\text{T} \,\left(T_\text{oven} - T \right) + \lambda \, h_\text{m} \phi\,S_\text{w}\,\left(\rho_\text{v} - \rho_\text{v,oven}\right),
\end{align}
and zero flux elsewhere. 
To enable comparison with previous implementations, we simulate a case by Ni~\cite{pm_ni_multiphase}, also published later by Datta~\cite{pm_datta_porous2007b}, where $\phi = 0.75$, $h_\text{T}= \SI{20} {W.m^{-2}.K^{-1}}$, $h_\text{m}= \SI{0.01}{m.s^{-1}}$ and $T_\text{oven} = \SI{450.15}{K}$. Simulations are initialized with $p=p_\text{amb}$, $S_\text{w}=0.5$ and $c_\text{v} = \SI{0.17}{\mol.m^{-3}}$.

We monitored the mesh quality and reduced the cell size until the mesh error remained significantly below one percent. Calculation of grid convergence and a possible implementation strategy in simulation software has been shown previously, see~\cite{diss_simconf19} for details.

\subsection{Discussion of results} 

\begin{figure}[hbt]
 \centering
 \scriptsize
\setlength\fheight{0.35\textheight}
\setlength\fwidth{0.465\textwidth}
%
%
\definecolor{mycolor1}{rgb}{0.00000,0.30588,0.54118}%
\definecolor{mycolor2}{rgb}{0.00000,0.61176,0.85490}%
\definecolor{mycolor3}{rgb}{0.65098,0.00000,0.51765}%
\begin{tikzpicture}

\begin{axis}[%
separate axis lines,
every outer x axis line/.append style={white!15!black},
every x tick label/.append style={font=\color{white!15!black}},
every x tick/.append style={white!15!black},
xmin=0,
xmax=0.01,
xtick={0,0.002,0.004,0.006,0.008,0.01},
xlabel style={font=\color{white!15!black}},
xlabel={Position  $y$ / m},
every outer y axis line/.append style={white!15!black},
every y tick label/.append style={font=\color{white!15!black}},
every y tick/.append style={white!15!black},
ymin=330,
ymax=410,
ylabel style={font=\color{white!15!black}},
ylabel={Temperature $T$ / K},
axis background/.style={fill=white},
title style={font=\bfseries},
axis on top,
legend style={legend cell align=left, align=left, draw=white!15!black},
legend style={at={(0.98,0.32)},anchor=south east},
width = \fwidth, 
 height = \fheight 
]
\addplot [color=black, line width=1.0pt, only marks, mark size=2.0pt, mark=+, mark options={solid, black}]
  table[]{tiks/datta1d/lowm-verif-noconv-T-1.tsv};
\addlegendentry{$T_\text{ref}$ at \SI{1200}{s}}

\addplot [color=black, line width=1.0pt, only marks, mark size=2.0pt, mark=o, mark options={solid, black}]
  table[]{tiks/datta1d/lowm-verif-noconv-T-2.tsv};
\addlegendentry{$T_\text{ref}$ at \SI{2400}{s}}

\addplot [color=black, line width=1.0pt, only marks, mark size=2.0pt, mark=asterisk, mark options={solid, black}]
  table[]{tiks/datta1d/lowm-verif-noconv-T-3.tsv};
\addlegendentry{$T_\text{ref}$ at \SI{3600}{s}}

\addplot [color=mycolor1, line width=1.0pt]
  table[]{tiks/datta1d/lowm-verif-noconv-T-4.tsv};
\addlegendentry{$T$ at \SI{1200}{s}}

\addplot [color=mycolor2, line width=1.0pt]
  table[]{tiks/datta1d/lowm-verif-noconv-T-5.tsv};
\addlegendentry{$T$ at \SI{2400}{s}}

\addplot [color=tud9c, line width=1.0pt]
  table[]{tiks/datta1d/lowm-verif-noconv-T-6.tsv};
\addlegendentry{$T$ at \SI{3600}{s}}

\end{axis}

\begin{axis}[%
every outer x axis line/.append style={white!15!black},
every x tick label/.append style={font=\color{white!15!black}},
every x tick/.append style={white!15!black},
xmin=0,
xmax=1,
xtick={0,0.1,0.2,0.3,0.4,0.5,0.6,0.7,0.8,0.9,1},
every outer y axis line/.append style={white!15!black},
every y tick label/.append style={font=\color{white!15!black}},
every y tick/.append style={white!15!black},
ymin=0,
ymax=1,
ytick={0,0.1,0.2,0.3,0.4,0.5,0.6,0.7,0.8,0.9,1},
axis line style={only marks},
ticks=none,
axis x line*=bottom,
axis y line*=left,
axis on top,
legend style={legend cell align=left, align=left, draw=white!15!black},
width = \fwidth, 
 height = \fheight 
]
\node[fill=white, below right, align=left, draw=black,anchor=north east]
at (rel axis cs:0.98,0.98) {MAPE = \SI{0.5}{\%} RMSE = \SI{1.9}{K} at \SI{1200}{s}\\MAPE = \SI{0.6}{\%} RMSE = \SI{2.2}{K}  at \SI{2400}{s}\\MAPE = \SI{0.6}{\%} RMSE = \SI{2.2}{K} at \SI{3600}{s}};
\end{axis}
\end{tikzpicture}%
\setlength\fheight{0.35\textheight}
\setlength\fwidth{0.52\textwidth}
%
%
\definecolor{mycolor1}{rgb}{0.00000,0.30588,0.54118}%
\definecolor{mycolor2}{rgb}{0.00000,0.61176,0.85490}%
\definecolor{mycolor3}{rgb}{0.65098,0.00000,0.51765}%
\begin{tikzpicture}

\begin{axis}[%
separate axis lines,
every outer x axis line/.append style={white!15!black},
every x tick label/.append style={font=\color{white!15!black}},
every x tick/.append style={white!15!black},
xmin=0,
xmax=0.01,
xtick={0,0.002,0.004,0.006,0.008,0.01},
xlabel style={font=\color{white!15!black}},
xlabel={Position  $y$ / m},
every outer y axis line/.append style={white!15!black},
every y tick label/.append style={font=\color{white!15!black}},
every y tick/.append style={white!15!black},
ymin=0,
ymax=0.66,
ytick={0,0.1,0.2,0.3,0.4,0.5,0.6},
ylabel style={font=\color{white!15!black}},
ylabel={$\text{Water Saturation } _\text{s}\text{ / -}$},
axis background/.style={fill=white},
title style={font=\bfseries},
axis on top,
legend style={legend cell align=left, align=left, draw=white!15!black},
legend style={at={(0.98,0.02)},anchor=south east},
width = \fwidth, 
 height = \fheight 
]
\addplot [color=black, line width=1.0pt, only marks, mark size=2.0pt, mark=+, mark options={solid, black}]
  table[]{tiks/datta1d/lowm-verif-noconv-Sl-1.tsv};
\addlegendentry{$S_\text{w,ref}\text{ at \SI{1200}{s}}$}

\addplot [color=black, line width=1.0pt, only marks, mark size=2.0pt, mark=o, mark options={solid, black}]
  table[]{tiks/datta1d/lowm-verif-noconv-Sl-2.tsv};
\addlegendentry{$S_\text{w,ref}\text{ at \SI{2400}{s}}$}

\addplot [color=black, line width=1.0pt, only marks, mark size=2.0pt, mark=asterisk, mark options={solid, black}]
  table[]{tiks/datta1d/lowm-verif-noconv-Sl-3.tsv};
\addlegendentry{$S_\text{w,ref}\text{ at \SI{3600}{s}}$}

\addplot [color=mycolor1, line width=1.0pt]
  table[]{tiks/datta1d/lowm-verif-noconv-Sl-4.tsv};
\addlegendentry{$S_\text{w}\text{ at \SI{1200}{s}}$}

\addplot [color=mycolor2, line width=1.0pt]
  table[]{tiks/datta1d/lowm-verif-noconv-Sl-5.tsv};
\addlegendentry{$S_\text{w}\text{ at \SI{2400}{s}}$}

\addplot [color=tud9c, line width=1.0pt]
  table[]{tiks/datta1d/lowm-verif-noconv-Sl-6.tsv};
\addlegendentry{$S_\text{w}\text{ at \SI{3600}{s}}$}

\end{axis}

\begin{axis}[%
every outer x axis line/.append style={white!15!black},
every x tick label/.append style={font=\color{white!15!black}},
every x tick/.append style={white!15!black},
xmin=0,
xmax=1,
xtick={0,0.1,0.2,0.3,0.4,0.5,0.6,0.7,0.8,0.9,1},
every outer y axis line/.append style={white!15!black},
every y tick label/.append style={font=\color{white!15!black}},
every y tick/.append style={white!15!black},
ymin=0,
ymax=1,
ytick={0,0.1,0.2,0.3,0.4,0.5,0.6,0.7,0.8,0.9,1},
axis line style={only marks},
ticks=none,
axis x line*=bottom,
axis y line*=left,
axis on top,
legend style={legend cell align=left, align=left, draw=white!15!black},
width = \fwidth, 
 height = \fheight 
]
\node[fill=white, below right, align=left, draw=black]
at (rel axis cs:0.02,0.98) {MAPE = \SI{1.2}{\%} RMSE = \num{4.8e-03} at \SI{1200}{s}\\MAPE = \SI{2.4}{\%} RMSE = \num{7.7e-03} at \SI{2400}{s}\\MAPE = \SI{3.6}{\%} RMSE = \num{9.8e-03}  at \SI{3600}{s}};
\end{axis}
\end{tikzpicture}%
\setlength\fheight{0.3\textheight}
\setlength\fwidth{0.62\textwidth}
%
%
\definecolor{mycolor1}{rgb}{0.00000,0.30588,0.54118}%
\definecolor{mycolor2}{rgb}{0.00000,0.61176,0.85490}%
\definecolor{mycolor3}{rgb}{0.65098,0.00000,0.51765}%
\begin{tikzpicture}

\begin{axis}[%
separate axis lines,
every outer x axis line/.append style={white!15!black},
every x tick label/.append style={font=\color{white!15!black}},
every x tick/.append style={white!15!black},
xmin=-0.0015,
xmax=0.01,
xtick={0,0.002,0.004,0.006,0.008,0.01},
xlabel style={font=\color{white!15!black}},
xlabel={Position  $y$ / m},
every outer y axis line/.append style={white!15!black},
every y tick label/.append style={font=\color{white!15!black}},
every y tick/.append style={white!15!black},
ymin=0,
ymax=490,
ytick={0,50,100,150,200,250,300,350,400,450},
ylabel style={font=\color{white!15!black}},
ylabel={Relative Pressure  $p$ / Pa},
axis background/.style={fill=white},
title style={font=\bfseries},
axis on top,
legend style={at={(0.02,0.98)}, anchor=north west, legend cell align=left, align=left, draw=white!15!black},
width = \fwidth, 
 height = \fheight 
]
\addplot [color=black, line width=1.0pt, only marks, mark size=2.0pt, mark=+, mark options={solid, black}]
  table[]{tiks/datta1d/lowm-verif-noconv-p-1.tsv};
\addlegendentry{$p_\text{ref}$ at \SI{1200}{s}}

\addplot [color=black, line width=1.0pt, only marks, mark size=2.0pt, mark=o, mark options={solid, black}]
  table[]{tiks/datta1d/lowm-verif-noconv-p-2.tsv};
\addlegendentry{$p_\text{ref}$ at \SI{2400}{s}}

\addplot [color=black, line width=1.0pt, only marks, mark size=2.0pt, mark=asterisk, mark options={solid, black}]
  table[]{tiks/datta1d/lowm-verif-noconv-p-3.tsv};
\addlegendentry{$p_\text{ref}$ at \SI{3600}{s}}

\addplot [color=mycolor1, line width=1.0pt]
  table[]{tiks/datta1d/lowm-verif-noconv-p-4.tsv};
\addlegendentry{$p$ at \SI{1200}{s} }

\addplot [color=mycolor2, line width=1.0pt]
  table[]{tiks/datta1d/lowm-verif-noconv-p-5.tsv};
\addlegendentry{$p$ at \SI{2400}{s}}

\addplot [color=tud9c, line width=1.0pt]
  table[]{tiks/datta1d/lowm-verif-noconv-p-6.tsv};
\addlegendentry{$p$ at \SI{3600}{s}}

\end{axis}

\begin{axis}[%
every outer x axis line/.append style={white!15!black},
every x tick label/.append style={font=\color{white!15!black}},
every x tick/.append style={white!15!black},
xmin=0,
xmax=1,
xtick={0,0.1,0.2,0.3,0.4,0.5,0.6,0.7,0.8,0.9,1},
every outer y axis line/.append style={white!15!black},
every y tick label/.append style={font=\color{white!15!black}},
every y tick/.append style={white!15!black},
ymin=0,
ymax=1,
ytick={0,0.1,0.2,0.3,0.4,0.5,0.6,0.7,0.8,0.9,1},
axis line style={only marks},
ticks=none,
axis x line*=bottom,
axis y line*=left,
axis on top,
legend style={legend cell align=left, align=left, draw=white!15!black},
width = \fwidth, 
 height = \fheight 
]
\node[fill=white, below right, align=left, draw=black,anchor=south east]
at (rel axis cs:0.98,0.025) {MAPE = \SI{6.2}{\%} RMSE = \SI{11.6}{\pascal} at \SI{1200}{s}\\MAPE = \SI{1.1}{\%} RMSE = \SI{3.5}{\pascal} at \SI{2400}{s}\\MAPE = \SI{0.7}{\%} RMSE = \SI{2.8}{\pascal} at \SI{3600}{s}};
\end{axis}
\end{tikzpicture}%
\setlength\fheight{0.3\textheight}
\setlength\fwidth{0.37\textwidth}
%
%
\definecolor{mycolor1}{rgb}{0.00000,0.30588,0.54118}%
\begin{tikzpicture}

\begin{axis}[%
separate axis lines,
every outer x axis line/.append style={white!15!black},
every x tick label/.append style={font=\color{white!15!black}},
every x tick label/.append style={/pgf/number format/set thousands separator={\,}},
every x tick/.append style={white!15!black},
xmin=-5,
xmax=3600,
xtick={0,1000,2000,3000,4000},
xlabel style={font=\color{white!15!black}},
xlabel={Time  $t$ / s},
every outer y axis line/.append style={white!15!black},
every y tick label/.append style={font=\color{white!15!black}},
every y tick/.append style={white!15!black},
ymin=0.0002,
ymax=0.0007,
ytick={0,0.0001,0.0002,0.0003,0.0004,0.0005,0.0006,0.0007},
ylabel style={font=\color{white!15!black}},
ylabel={$\text{Evap. } \dot{m}_{\text{loss}}\text{ / } \si{ kg.m^{-2}.s^{-1} }$},
axis background/.style={fill=white},
title style={font=\bfseries},
axis on top,
legend style={legend cell align=left, align=left, draw=white!15!black},
width = \fwidth, 
 height = \fheight 
]
\addplot [color=black, line width=1.0pt, only marks, mark size=2.0pt, mark=+, mark options={solid, black},each nth point={2}]
  table[]{tiks/datta1d/lowm-verif-noconv-mloss-1.tsv};
\addlegendentry{$\dot{m}_{\text{loss,ref}}$}

\addplot [color=mycolor1, line width=1.0pt]
  table[]{tiks/datta1d/lowm-verif-noconv-mloss-2.tsv};
\addlegendentry{${m}_{\text{loss}}$}

\end{axis}

\begin{axis}[%
every outer x axis line/.append style={white!15!black},
every x tick label/.append style={font=\color{white!15!black}},
every x tick/.append style={white!15!black},
xmin=0,
xmax=1,
xtick={0,0.1,0.2,0.3,0.4,0.5,0.6,0.7,0.8,0.9,1},
every outer y axis line/.append style={white!15!black},
every y tick label/.append style={font=\color{white!15!black}},
every y tick/.append style={white!15!black},
ymin=0,
ymax=1,
ytick={0,0.1,0.2,0.3,0.4,0.5,0.6,0.7,0.8,0.9,1},
axis line style={only marks},
ticks=none,
axis x line*=bottom,
axis y line*=left,
axis on top,
legend style={legend cell align=left, align=left, draw=white!15!black},
width = \fwidth, 
 height = \fheight 
]
\node[fill=white, below right, align=left, draw=black, anchor=south east]
at (rel axis cs:0.98,0.02) {MAPE = \SI{4.0}{\%} \\ RMSE = \SI{3.3e-05}{kg.m^{-2}.s^{-1}} } ;
\end{axis}
\end{tikzpicture}%
 \caption{Spatial comparison of implemented porous media model with reference data~\cite{pm_ni_multiphase}.}
 \label{fig:d1d-valid-x}
\end{figure}
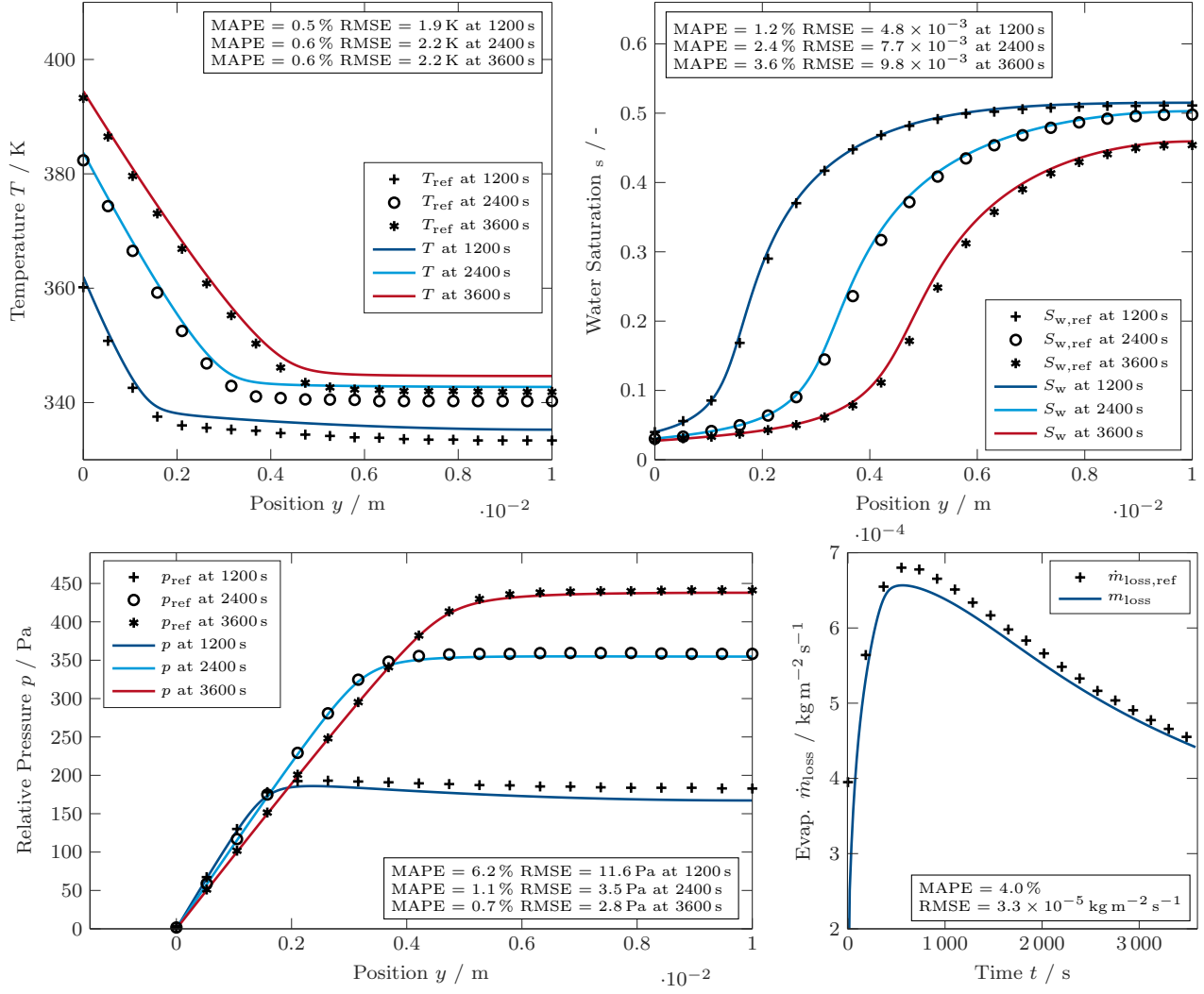

The Mean Absolute Percentage Error (MAPE) and Root Mean Square Error~(RMSE) are chosen to quantify errors in a robust manner:
\begin{align}
\text{MAPE} = \frac{1}{N} \sum_{i=1}^N \frac{|R_i - F_i|}{\frac{1}{2} |R_i + F_i|}\times 100\,\%\,, \quad \text{RMSE} = \sqrt{ \frac{1}{N} \sum_{i=1}^N (R_i - F_i)^2}\,,
\end{align}
where $R_i$ are realisations of the reference and $F_i$ are forecasts by our simulation at sampled points in time $t_i$. 
The evaluation of the implemented model compared to the reference data of Ni~\cite{pm_ni_multiphase} is illustrated in Fig.~\ref{fig:d1d-valid-x}. 
The obtained results are in very good agreement with reference data, although the simulations have been originally performed on a custom code without the non-equilibrium evaporation formulation. Ni used a relatively coarse uniformly spaced grid with 41 elements and central differencing scheme. As it can be seen in Fig.~\ref{fig:d1d-valid-x}, the deviations grow constantly over time for temperatures (maximum MAPE = \SI{0.6}{\%}) and water saturations (maximum MAPE = \SI{3.6}{\%}).
We see too much progress of temperature and too few losses of water saturation. The influence of the non-equilibrium evaporation formulation has been investigated thoroughly. As $K_\text{evap} = \SI{1000}{s^{-1}}$ enforces the system to an almost instantaneous change to equilibrium, we see no significant change in results when further increasing the value.
Remarkable is the difference of factor two in the evaporative mass loss at initialisation of the simulation. Here, the results of the reference and our model should be identical. Insufficient digits in the given mass transfer coefficient (MTC) are a possible error source. An MTC $h_\text{m} =\SI{0.0125}{m.s^{-1}}$ already shifts all values in the desired window. Here, we see a good example of how sensitive a food system reacts to fluctuations in heat transfer coefficients (HTC) and MTCs. As we will see in Sec.~\ref{sec:summary}, it is one of our aims to remove this dependency by treating the food process and adjacencies in a conjugate manner.

\section{NONLINEAR SYSTEM IDENTIFICATION}
\label{sec:dynrom}
As it was motivated in Sec.~\ref{sec:hybridtwin}, there is a need for substantial speed-up of the simulation model. In the research project framework, we opt for a non-intrusive ROM implemented in ANSYS Dynamic ROM Builder. It is a highly accurate data-driven method which does not rely on the physical model. This feature makes it universally applicable to various kinds of physics. It can be considered as a Non-linear AutoRegressive eXogenous (NARX) grey-box approach, based on a generalised ODE system ansatz with successively added additional degrees of freedom to cover nonlinearities and hysteresis effects~\cite{rom_twinbuilder_2020,rom_ljung_system,rom_nelles_nonlinear2021}. The parameter identification of the ODE system is accomplished with machine learning approaches~\cite{rom_twinbuilder_2020}.

In the following section, we discuss the impact of excitation signal synthesis on successful system identification. The system input of the Full Order Model (FOM) presented in Sec.~\ref{sec:porousmedia} is the oven temperature. The output of our Single Input Single Output (SISO) case is the core temperature history $T_\text{core} (t , y=\SI{0.01}{m})$. Temperature is the obvious first choice for a test setup, as nonlinearities and progress dependent effects occur. 

\subsection{Excitation signal and training data synthesis}
\begin{figure}[hbtp]
 \centering
 \setlength\fheight{0.23\textheight}
\setlength\fwidth{0.495\textwidth}
%
%
\definecolor{mycolor1}{rgb}{0.75294,0.75294,0.75294}%
\definecolor{mycolor2}{rgb}{0.50196,0.50196,0.50196}%
\begin{tikzpicture}

\begin{axis}[%
xmin=-100,
xmax=9800,
xlabel style={font=\color{white!15!black}},
every x tick label/.append style={/pgf/number format/set thousands separator={\,}},
xlabel={Time $t$ / s},
ymin=279.149364077244,
ymax=445,
ylabel style={font=\color{white!15!black}},
ylabel={Temperature $T$ / K},
axis background/.style={fill=white},
legend style={at={(0.5,1.2)},anchor=north, legend columns=3, legend cell align=left, align=left, draw=white!15!black},
width = \fwidth, 
 height = \fheight 
]
\addplot [color=black, line width=1.2pt]
  table[]{tiks/dyn-ecco/sig3-3-e-1.tsv};
\addlegendentry{$T_\text{core}$ FOM}

\addplot [color=mycolor1, dashed, line width=1.2pt]
  table[]{tiks/dyn-ecco/sig3-3-e-2.tsv};
\addlegendentry{$T_\text{core}$ ROM}

\addplot [color=tud9c, line width=1.2pt]
  table[]{tiks/dyn-ecco/sig3-3-e-3.tsv};
\addlegendentry{$\text{T}_{\text{oven}}$}

\end{axis}

\begin{axis}[%
xmin=0,
xmax=1,
ymin=0,
ymax=1,
axis line style={only marks},
ticks=none,
axis x line*=bottom,
axis y line*=left,
legend style={legend cell align=left, align=left, draw=white!15!black},
width = \fwidth, 
 height = \fheight 
]
\node[fill=white, below right, align=left, draw=black]
at (rel axis cs:0.025,0.9)  {\textbf{Case A}\\ MAPE = \SI{0.0}{\%}\\ RMSE = \SI{0.1}{K}\\  MAX = \SI{0.4}{K}};
\end{axis}
\end{tikzpicture}%
\hfill
%
%
\definecolor{mycolor1}{rgb}{0.75294,0.75294,0.75294}%
\definecolor{mycolor2}{rgb}{0.50196,0.50196,0.50196}%
\begin{tikzpicture}

\begin{axis}[%
xmin=0,
xmax=9800,
xlabel style={font=\color{white!15!black}},
xlabel={Time  $t$ / s},
every x tick label/.append style={/pgf/number format/set thousands separator={\,}},
ymin=279.024836527793,
ymax=424,
ylabel style={font=\color{white!15!black}},
ylabel={Temperature $T$ / K},
axis background/.style={fill=white},
legend style={at={(0.5,1.2)},anchor=north, legend columns=3, legend cell align=left, align=left, draw=white!15!black},
width = \fwidth, 
 height = \fheight 
]
\addplot [color=black, line width=1.2pt]
  table[]{tiks/dyn-ecco/sig3-2-e-1.tsv};

\addplot [color=mycolor1, dashed, line width=1.2pt]
  table[]{tiks/dyn-ecco/sig3-2-e-2.tsv};

\addplot [color=tud9c, line width=1.2pt]
  table[]{tiks/dyn-ecco/sig3-2-e-3.tsv};

\end{axis}

\begin{axis}[%
xmin=0,
xmax=1,
ymin=0,
ymax=1,
axis line style={only marks},
ticks=none,
axis x line*=bottom,
axis y line*=left,
legend style={legend cell align=left, align=left, draw=white!15!black},
width = \fwidth, 
 height = \fheight 
]
\node[fill=white, below right, align=left, draw=black, anchor=north west]
at (rel axis cs:0.025,0.9)  {\textbf{Case B}\\ MAPE = \SI{0.0}{\%} \\RMSE = \SI{0.1}{K} \\ MAX = \SI{0.3}{K}};
\end{axis}
\end{tikzpicture}%
%
%
\definecolor{mycolor1}{rgb}{0.75294,0.75294,0.75294}%
\definecolor{mycolor2}{rgb}{0.50196,0.50196,0.50196}%
\begin{tikzpicture}

\begin{axis}[%
xmin=0,
xmax=9800,
xlabel style={font=\color{white!15!black}},
every x tick label/.append style={/pgf/number format/set thousands separator={\,}},
xlabel={Time $t$ / s},
ymin=279.104072105527,
ymax=413,
ylabel style={font=\color{white!15!black}},
ylabel={Temperature $T$ / K},
axis background/.style={fill=white},
legend style={legend cell align=left, align=left, draw=white!15!black},
width = \fwidth, 
 height = \fheight 
]
\addplot [color=black, line width=1.2pt]
  table[]{tiks/dyn-ecco/sig3-4-e-1.tsv};

\addplot [color=mycolor1, dashed, line width=1.2pt]
  table[]{tiks/dyn-ecco/sig3-4-e-2.tsv};

\addplot [color=tud9c, line width=1.2pt]
  table[]{tiks/dyn-ecco/sig3-4-e-3.tsv};

\end{axis}

\begin{axis}[%
xmin=0,
xmax=1,
ymin=0,
ymax=1,
axis line style={only marks},
ticks=none,
axis x line*=bottom,
axis y line*=left,
legend style={legend cell align=left, align=left, draw=white!15!black},
width = \fwidth, 
 height = \fheight 
]
\node[fill=white, below right, align=left, draw=black]
at (rel axis cs:0.025,0.9) {\textbf{Case C}\\ MAPE = \SI{0.0}{\%}\\ RMSE = \SI{0.1}{K} \\ MAX = \SI{0.5}{K}};
\end{axis}
\end{tikzpicture}%
 \hfill
%
%
\definecolor{mycolor1}{rgb}{0.75294,0.75294,0.75294}%
\definecolor{mycolor2}{rgb}{0.50196,0.50196,0.50196}%
\begin{tikzpicture}

\begin{axis}[%
xmin=-100,
xmax=9800,
xlabel style={font=\color{white!15!black}},
every x tick label/.append style={/pgf/number format/set thousands separator={\,}},
xlabel={Time $t$ / s},
ymin=279.15,
ymax=455,
ylabel style={font=\color{white!15!black}},
ylabel={Temperature $T$ / K},
axis background/.style={fill=white},
legend style={legend cell align=left, align=left, draw=white!15!black},
width = \fwidth, 
 height = \fheight 
]
\addplot [color=black, line width=1.2pt]
  table[]{tiks/dyn-ecco/sig3-7-e-1.tsv};

\addplot [color=mycolor1, dashed, line width=1.2pt]
  table[]{tiks/dyn-ecco/sig3-7-e-2.tsv};

\addplot [color=tud9c, line width=1.2pt]
  table[]{tiks/dyn-ecco/sig3-7-e-3.tsv};

\end{axis}

\begin{axis}[%
xmin=0,
xmax=1,
ymin=0,
ymax=1,
axis line style={only marks},
ticks=none,
axis x line*=bottom,
axis y line*=left,
legend style={legend cell align=left, align=left, draw=white!15!black},
width = \fwidth, 
 height = \fheight 
]
\node[fill=white, below right, align=left, draw=black]
at (rel axis cs:0.025,0.9)  {\textbf{Case D}\\ MAPE = \SI{0.0}{\%}\\ RMSE = \SI{0.1}{K}\\  MAX = \SI{0.4}{K}};
\end{axis}
\end{tikzpicture}%
 \caption{Trained cases A to D.}
 \label{fig:dynrom-d1d-trained}
\end{figure}
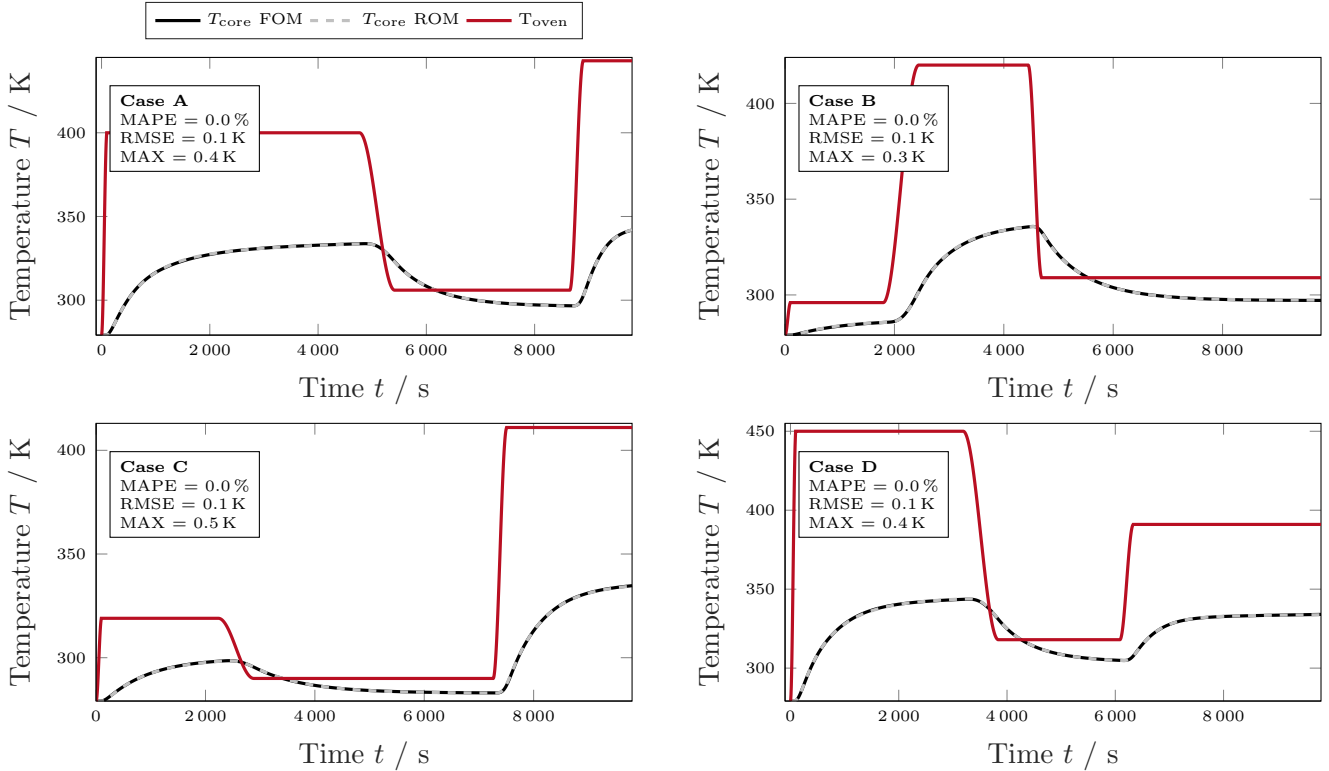

The successful identification of non-linear dynamic models strongly depends on the provided training data.
We know from fundamental linear dynamics: Its step response can fully characterise a linear model.
On the contrary, the input and output spaces of non-linear models have to be covered systematically.
Nelles~\cite{rom_nelles_nonlinear2021} recommends applying an Amplitude modulated Pseudo-Random Binary Signal (APBRS).
Due to the input signal’s rate constraints -- e.g. the oven temperature cannot change instantly~-- sined transitions of different frequencies model different heating rates. A minimum hold time of \SI{500}{s} for the input allows the process to fulfil its full response.
The remaining settings are: $T \in [\SI{280}{K},\SI{450}{K}] , f_\text{sine} \in [\SI{0.0017 }{Hz},\SI{0.0017 }{Hz}]$.
We can identify a ROM of the cooking process with only four randomly chosen training cases, compare Fig.~\ref{fig:dynrom-d1d-trained}. 
The RMSE for the training cases remains at \SI{0.1}{K}.

\subsection{Performance analysis} \label{sec:performanceanalysis}
\begin{figure}[htbp]
 \centering
 \setlength\fheight{0.23\textheight}
\setlength\fwidth{0.495\textwidth}
%
%
\definecolor{mycolor1}{rgb}{0.75294,0.75294,0.75294}%
\definecolor{mycolor2}{rgb}{0.50196,0.50196,0.50196}%
\begin{tikzpicture}

\begin{axis}[%
xmin=0,
xmax=12500,
xlabel style={font=\color{white!15!black}},
xlabel={Time $t$ / s},
ymin=279.070084824962,
ymax=437,
ylabel style={font=\color{white!15!black}},
ylabel={Temperature $T$ / K},
axis background/.style={fill=white},
legend style={at={(0.5,1.2)},anchor=north, legend columns=3, legend cell align=left, align=left, draw=white!15!black},
width = \fwidth, 
 height = \fheight 
]
\addplot [color=black, line width=1.2pt]
  table[]{tiks/dyn-ecco/sig4-1-e-1.tsv};
\addlegendentry{$T_\text{core}$ FOM}

\addplot [color=mycolor1, dashed, line width=1.2pt]
  table[]{tiks/dyn-ecco/sig4-1-e-2.tsv};
\addlegendentry{$T_\text{core}$ ROM}

\addplot [color=tud9c, line width=1.2pt]
  table[]{tiks/dyn-ecco/sig4-1-e-3.tsv};
\addlegendentry{$\text{T}_{\text{oven}}$}

\end{axis}

\begin{axis}[%
xmin=0,
xmax=1,
ymin=0,
ymax=1,
axis line style={only marks},
ticks=none,
axis x line*=bottom,
axis y line*=left,
legend style={legend cell align=left, align=left, draw=white!15!black},
width = \fwidth, 
 height = \fheight 
]
\node[fill=white, below right, align=left, draw=black, anchor=north west]
at (rel axis cs:0.025,0.9)  {\textbf{Case E} \\ MAPE = \SI{0.1}{\%}\\RMSE = \SI{0.5}{K} \\MAX = \SI{0.8}{K}};
\end{axis}
\end{tikzpicture}%
 \hfill
%
%
\definecolor{mycolor1}{rgb}{0.75294,0.75294,0.75294}%
\definecolor{mycolor2}{rgb}{0.50196,0.50196,0.50196}%
\begin{tikzpicture}

\begin{axis}[%
xmin=-100,
xmax=9800,
xlabel style={font=\color{white!15!black}},
every x tick label/.append style={/pgf/number format/set thousands separator={\,}},
xlabel={Time $t$ / s},
ymin=279.149119703266,
ymax=395,
ytick={300,350,400},
ylabel style={font=\color{white!15!black}},
ylabel={Temperature $T$ / K},
axis background/.style={fill=white},
legend style={legend cell align=left, align=left, draw=white!15!black},
width = \fwidth, 
 height = \fheight 
]
\addplot [color=black, line width=1.2pt]
  table[]{tiks/dyn-ecco/sig3-9-e-1.tsv};

\addplot [color=mycolor1, dashed, line width=1.2pt]
  table[]{tiks/dyn-ecco/sig3-9-e-2.tsv};

\addplot [color=tud9c, line width=1.2pt]
  table[]{tiks/dyn-ecco/sig3-9-e-3.tsv};

\end{axis}

\begin{axis}[%
xmin=0,
xmax=1,
ymin=0,
ymax=1,
axis line style={only marks},
ticks=none,
axis x line*=bottom,
axis y line*=left,
legend style={legend cell align=left, align=left, draw=white!15!black},
width = \fwidth, 
 height = \fheight 
]
\node[fill=white, below right, align=left, draw=black]
at (rel axis cs:0.025,0.9) {\textbf{Case F} \\MAPE = \SI{0.2}{\%} \\RMSE = \SI{0.7}{K} \\ MAX = \SI{1.5}{K}};
\end{axis}
\end{tikzpicture}%
%
%
\definecolor{mycolor1}{rgb}{0.75294,0.75294,0.75294}%
\definecolor{mycolor2}{rgb}{0.50196,0.50196,0.50196}%
\begin{tikzpicture}

\begin{axis}[%
xmin=-100,
xmax=12500,
xlabel style={font=\color{white!15!black}},
xlabel={Time $t$ / s},
ymin=279.142409117884,
ymax=413,
ylabel style={font=\color{white!15!black}},
ylabel={Temperature $T$ / K},
axis background/.style={fill=white},
legend style={legend cell align=left, align=left, draw=white!15!black},
width = \fwidth, 
 height = \fheight 
]
\addplot [color=black, line width=1.2pt]
  table[]{tiks/dyn-ecco/sig4-9-e-1.tsv};

\addplot [color=mycolor1, dashed, line width=1.2pt]
  table[]{tiks/dyn-ecco/sig4-9-e-2.tsv};

\addplot [color=tud9c, line width=1.2pt]
  table[]{tiks/dyn-ecco/sig4-9-e-3.tsv};

\end{axis}

\begin{axis}[%
xmin=0,
xmax=1,
ymin=0,
ymax=1,
axis line style={only marks},
ticks=none,
axis x line*=bottom,
axis y line*=left,
legend style={legend cell align=left, align=left, draw=white!15!black},
width = \fwidth, 
 height = \fheight 
]
\node[fill=white, below right, align=left, draw=black, anchor=north west]
at (rel axis cs:0.025,0.98) {\textbf{Case G} \\MAPE = \SI{0.3}{\%}\\ RMSE = \SI{1.2}{K} \\ MAX = \SI{2.4}{K}};
\end{axis}
\end{tikzpicture}%
 \hfill
%
%
\definecolor{mycolor1}{rgb}{0.75294,0.75294,0.75294}%
\definecolor{mycolor2}{rgb}{0.50196,0.50196,0.50196}%
\begin{tikzpicture}

\begin{axis}[%
xmin=-100,
xmax=12500,
xlabel style={font=\color{white!15!black}},
xlabel={Time $t$ / s},
ymin=279.149405740573,
ymax=443,
ylabel style={font=\color{white!15!black}},
ylabel={Temperature $T$ / K},
axis background/.style={fill=white},
width = \fwidth, 
 height = \fheight 
]
\addplot [color=black, line width=1.2pt]
  table[]{tiks/dyn-ecco/sig4-2-e-1.tsv};

\addplot [color=mycolor1, dashed, line width=1.2pt]
  table[]{tiks/dyn-ecco/sig4-2-e-2.tsv};

\addplot [color=tud9c, line width=1.2pt]
  table[]{tiks/dyn-ecco/sig4-2-e-3.tsv};

\end{axis}

\begin{axis}[%
xmin=0,
xmax=1,
ymin=0,
ymax=1,
axis line style={only marks},
ticks=none,
axis x line*=bottom,
axis y line*=left,
legend style={legend cell align=left, align=left, draw=white!15!black},
width = \fwidth, 
 height = \fheight 
]
\node[fill=white, below right, align=left, draw=black, anchor=north west]
at (rel axis cs:0.025,0.98) {\textbf{Case H} \\MAPE = \SI{0.3}{\%}\\ RMSE = \SI{0.9}{K} \\ MAX = \SI{2.5}{K}};
\end{axis}
\end{tikzpicture}%
 \caption{Comparison of FOM and ROM for untrained cases E to H.}
 \label{fig:dynrom-d1d-eval}
\end{figure}
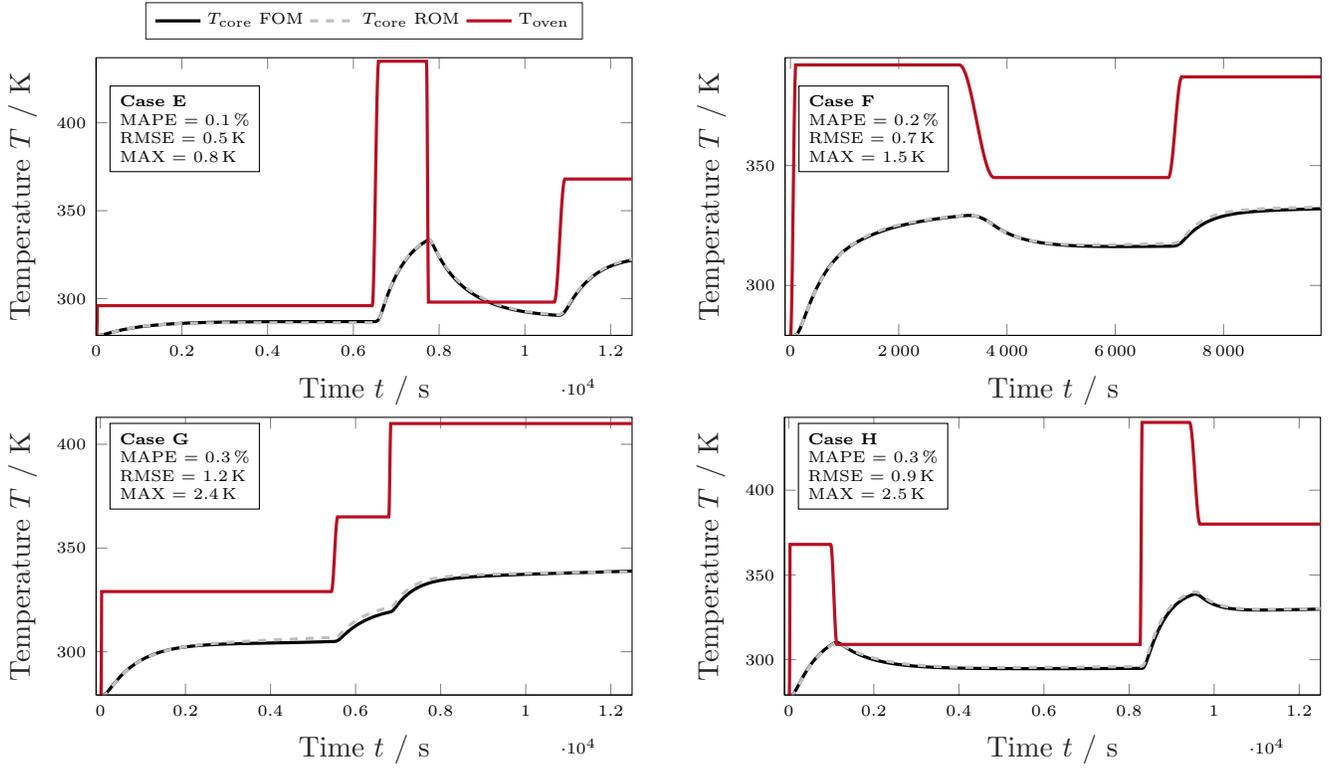

The ROM performance is evaluated with five randomly generated, sined APBRS excitation signals following the design rules outlined above.
Fig.~\ref{fig:dynrom-d1d-eval} illustrates that the errors remain significantly low with the MAPE not exceeding \SI{0.3}{\%} at any time.  
Case~G shows the overall maximum RMSE of \SI{1.2}{K}. Case~H is excited with high oven temperatures during the end of the process, where the overall maximum error of \SI{2.5}{K} occurs. The underlying hysteresis of cooking history appears to be more difficult to capture. The system’s effective thermal conductivity is strongly impacted by reduced water concentrations, compare progress of $S_\text{w}$ in Fig.~\ref{fig:d1d-valid-x} and composition of $k_\text{eff}$ in Eq.~\ref{eq:keff}. The limitation of temperatures due to evaporative losses for oven temperatures above \SI{373.15}{K} seems to be captured well in all cases. A clear dependency on different rates of change in the oven temperature cannot be distinguished.

\begin{figure}[htbp]
 \centering
\setlength\fheight{0.23\textheight}
\setlength\fwidth{0.33\textwidth}
%
%
\definecolor{mycolor1}{rgb}{0.75294,0.75294,0.75294}%
\definecolor{mycolor2}{rgb}{0.50196,0.50196,0.50196}%
\begin{tikzpicture}

\begin{axis}[%
xmin=-100,
xmax=12500,
xlabel style={font=\color{white!15!black}},
xlabel={Time  t / s},
ymin=260,
ymax=450,
ylabel style={font=\color{white!15!black}},
ylabel={Temperature T / K},
axis background/.style={fill=white},
title style={font=\bfseries},
legend style={legend cell align=left, align=left, draw=white!15!black},
width = \fwidth, 
 height = \fheight 
]
%

\addplot [color=tud9c, line width=1.2pt]
  table[]{tiks/dyn-ecco/sig4-7-C1-3.tsv};
\addlegendentry{$T_\text{oven}$}

\end{axis}

\begin{axis}[%
xmin=0,
xmax=1,
ymin=0,
ymax=1,
axis line style={only marks},
ticks=none,
axis x line*=bottom,
axis y line*=left,
legend style={legend cell align=left, align=left, draw=white!15!black},
width = \fwidth, 
 height = \fheight 
]
\end{axis}
\end{tikzpicture}%
\hfill
\setlength\fheight{0.23\textheight}
\setlength\fwidth{0.65\textwidth}
%
%
\definecolor{mycolor0}{rgb}{0.75294,0.75294,0.75294}%
\definecolor{mycolor1}{rgb}{0.00000,0.30588,0.54118}%
\definecolor{mycolor2}{rgb}{0.00000,0.61176,0.85490}%
\definecolor{mycolor3}{rgb}{0.65098,0.00000,0.51765}%
\begin{tikzpicture}

\begin{axis}[%
xmin=0,
xmax=12500,
xlabel style={font=\color{white!15!black}},
xlabel={Time  t / s},
ymin=279.149891181379,
ymax=350,
ylabel style={font=\color{white!15!black}},
ylabel={Temperature T / K},
axis background/.style={fill=white},
legend style={legend cell align=left, align=left, draw=white!15!black},
width = \fwidth, 
 height = \fheight 
]

\addplot [color=black, line width=1.2pt]
  table[]{tiks/dyn-ecco/sig4-7-1.tsv};
\addlegendentry{FOM}

\addplot [color=mycolor0, dashed, line width=1.2pt]
  table[]{tiks/dyn-ecco/sig4-7-2.tsv};
\addlegendentry{ROM}

\addplot [color=mycolor2, line width=1.2pt]
  table[]{tiks/dyn-ecco/sig4-7-C1-2.tsv};
\addlegendentry{Linear ROM}

\end{axis}

\begin{axis}[%
xmin=0,
xmax=1,
ymin=0,
ymax=1,
axis line style={only marks},
ticks=none,
axis x line*=bottom,
axis y line*=left,
legend style={legend cell align=left, align=left, draw=white!15!black},
width = \fwidth, 
 height = \fheight 
]
\node[fill=white, below right, align=left, draw=black,anchor=south east]
at (rel axis cs:0.98,0.02) {\textbf{Case I} \\MAPE = \SI{0.1}{\%} RMSE = \SI{0.4}{K}  MAX = \SI{0.9}{K}\\ MAPE = \SI{1.5}{\%} RMSE =  \SI{5.4}{K}  MAX = \SI{9.6}{K} (Linear ROM)};
\end{axis}
\end{tikzpicture}%
 \caption{Comparison of linear and non-linear ROM for case I.}
 \label{fig:rom_lin_vs_nonlin}
\end{figure}
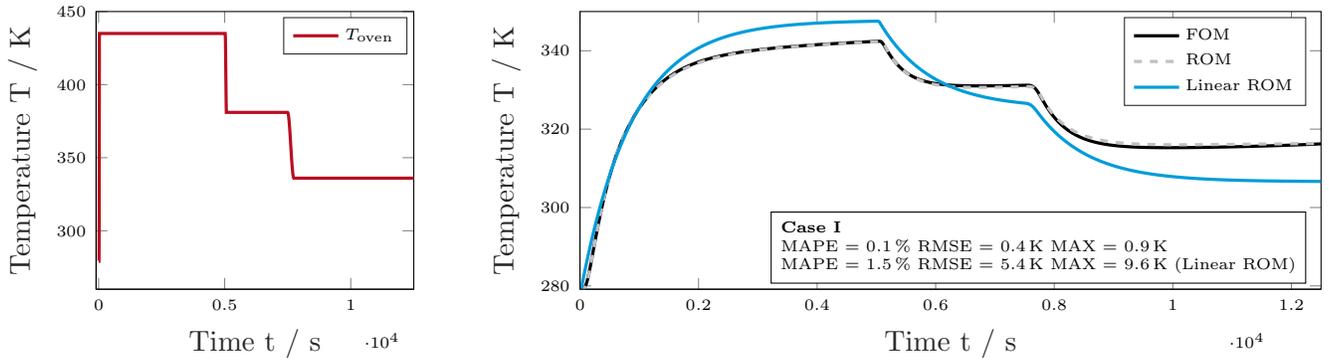

To better rank the non-linear ROM performance, a linear ROM is built with identical training cases. It cannot follow the temperature-dependent evaporation effects from the first excitation step and onwards, compare Fig.~\ref{fig:rom_lin_vs_nonlin}. 
There is no trend of just over or under-predicting food temperatures. At different oven temperatures, we see the non-linear model following the different system dynamics in each step with a maximum error of \SI{0.9}{K}, which is one order of magnitude lower than for the linear model.

The obtained speed-up of the ROM is calculated by comparing three cases, see Tab.~\ref{tab:predictionperformance}. The data in the first row origins from a previous study~\cite{diss_simconf19}. It symbolises the time for the 3D simulation of a cooking device. It is a conjugate heat model with a Discrete Transfer radiation model. As we included no food-specific models, we expect even more than the required two hours CPU time for a \SI{120}{s} prediction for a fully coupled model. Secondly, we see the calculation times of \SI{2830}{s} for the 1D simplified cooking model that we introduced in Sec.~\ref{sec:porousmedia}.

The extracted non-linear ROM was exported as an FMU and executed in an open-loop setup in ANSYS Twin Builder. The time to predict \SI{10000}{s} of real-time was found to be \SI{0.3}{s} which equals to a speedup of factor \num{33 333} compared to \textit{real-time} simulations. To put it differently, this means 556 predictions \textit{per minute} of the following hour.
\newcommand{\specialcell}[2][l]{%
\begin{tabular}[#1]{@{}c@{}}#2\end{tabular}}
\begin{table}[hbtp]\centering\footnotesize
\ra{1.1}
\caption{Comparison of prediction time performance of CFD versus ROM.}
\vspace*{1em}
\begin{tabular}{@{}lrrllrr@{}} \toprule
 Type        & Real-time & \specialcell{CPU \\time} & \specialcell{CPU load on \\ 16 core Xeon E5 3 GHz }    & Problem info & \specialcell{Time per \\1h prediction} & \\
 \midrule
CFD      & \SI{120}{s}  & \SI{7200}{s} & Parallelized, full load & 500k DOFs, CHT only & \SI{216000}{s}   & \\
CFD 		& \SI{10250}{s} & \SI{2830}{s} & Parallelized, full load & 1876 DOFs & \SI{994} {s} \\
DynROM & \SI{10000}{s}  & \SI{0.3}{s} & Serial, no noticable CPU load  & FMU, SISO & \SI{0.1}{s}  & \\
\bottomrule  
 \label{tab:predictionperformance}        &   &  &      &   &   & 
\end{tabular}
\end{table}

\section{SUMMARY}
\label{sec:summary}
This work motivates the need for autonomous cooking processes and proposes the application of DT methodology.
We present a successful implementation of a porous media model to simulate food processes. It serves to simulate non-measurable quantities of interest. 
We follow a hybrid approach: Full order simulations of multi-physical cooking processes form the basis for subsequent data-driven system identification. The ROM can reproduce temperature histories of up to \SI{12500}{s} with an average RMSE of less than one Kelvin. The maximum observed deviation of the non-linear ROM is \SI{2.4}{K}, although evaporation and hysteresis effects introduce strong non-linear model behaviour. The RMSE remains one order of magnitude lower than for linear ROMs. 
 
Based on literature research and to our best knowledge, this work introduces the first DT based framework for autonomous cooking processes that fulfils real-time data provision.
The recent review of Verboven~\cite{dt_verboven_digital2020} marks the dawn of DTs in the Food Science community. Their realisations of DTs are only post-processing steps included in a vision of cloud-based services.
On the contrary, our approach exceeds the minimum requirement of real-time simulations with a large margin. The ROM provides more than 550 predictions of one hour real-time per minute. Faster-than-real-time simulations of DTs provide valuable insights into the current and multiple possible future states. This potential can be exploited best in optimal control algorithms. 
Due to drastically reduced computational costs, the method is available for live on-board autonomous operations. 
The rise of high performance or cloud computing is assumed to be too slow to become a feasible alternative for 3D coupled multi-physical simulations on food processing level in the foreseeable future. 
In general, autonomous processes based on DTs can be regarded as an alternative to the current Internet of Things (IoT) trend of connected devices. 
There is no need for communication hardware, security protocols or complex sensor arrays to capture state information. DTs provide all relevant data offline and on-board.

\paragraph{Outlook: The need for coupled treatment of surrounding process and food} 

\begin{figure}[hbtp]
 \centering
 \captionsetup[subfigure]{justification=centering}
\centering
\begin{subfigure}[c]{0.495\textwidth}
 \includegraphics[width=\textwidth]{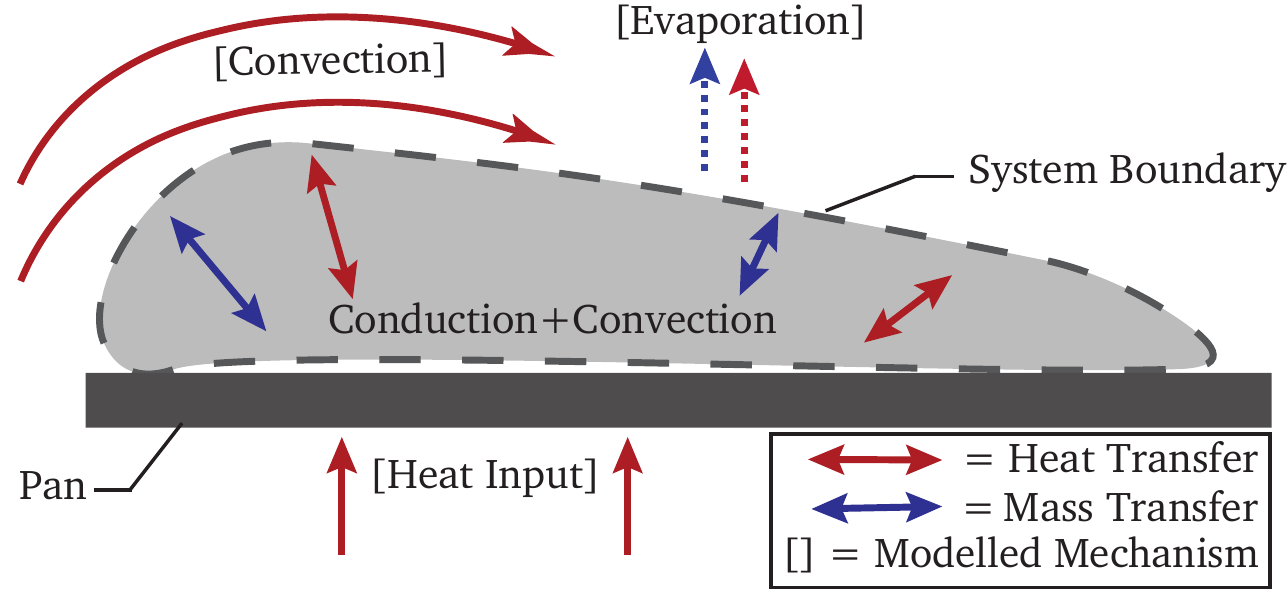}
\subcaption{Food-only modelling.\label{fig:vof-motivation-a}}
\end{subfigure}
 \captionsetup[subfigure]{justification=centering}
\centering
\begin{subfigure}[c]{0.495\textwidth}
\includegraphics[width=\textwidth]{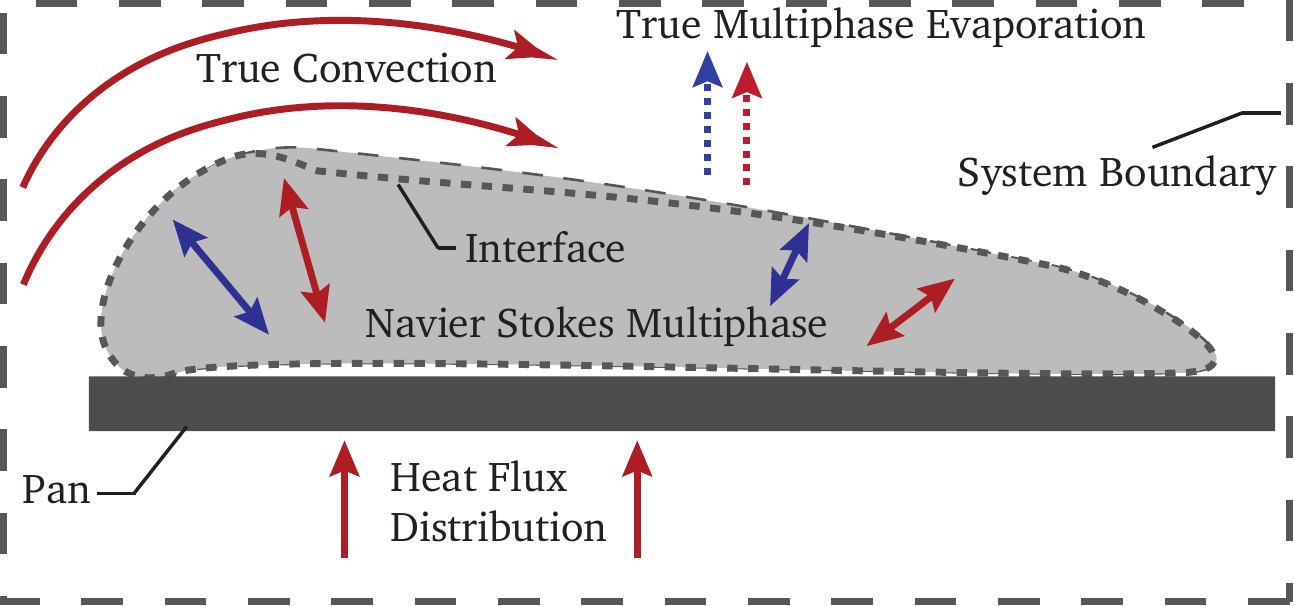}
\subcaption{Conjugate modelling of food and adjacencies.\label{fig:vof-motivation-b}}
\end{subfigure}
 \caption{Comparison of transfer coefficient modelling and coupled treatment.}
 \label{fig:vof-motivation}
\end{figure}

Future works should address the coupled treatment of the food processing device (often modelled as thermal fluid-structure interaction) and the cooking process itself (heat and mass transfer in porous media). 
The coupling with adjacencies is vital for accuracy due to numerous reasons:
In food science, a particular focus lies in the modelling of only food. As illustrated in Fig.~\ref{fig:vof-motivation-a}, the food’s surface forms the system boundary. 
This implies using boundary conditions to model convection, evaporation or radiation. HTCs and MTCs represent the heat and mass exchange with the cooking process. However, these are seldom constant in time and space. 
Halder~\cite{pm_halder_surface12}, for instance, concludes: \enquote{[...] a lumped heat and mass transfer coefficient, which includes the effects of both diffusive and convective flow, [...] is not expected to be constant over time [...]. [The HTC] varies significantly along the surface}. On top of inherently transient fluctuations due to buoyancy, we can see a position dependence in cooking devices, e.g. due to flow pattern induced by the topology. 

Moreover, there are food specific impacts: Ibarra~\cite{ir_ibarra_cooked2000} for example documented the progress dependent emissivity of chicken fillets. Almeida~\cite{ir_almeida_measurement2006} determined that the spectral absorbance of potato is not constant over frequency.

The coupling of process and product becomes evident with the following example: The heat flux of an oven is not turned off by disabling the heating system, as passive heating persists due to the walls’ presence. 
Therefore, future work should focus on the coupled interaction of device and food. A possible implementation is illustrated in Fig.~\ref{fig:vof-motivation-b}. One could envision a coupled multi-phase model (e.g. Volume Of Fluid method) on one connected domain that also accounts for the food-specific processes.

\section*{ACKNOWLEDGEMENTS}
The work of Maximilian Kannapinn is supported by the Graduate School CE within the Centre for Computational Engineering at Technische Universität Darmstadt. 

\appendix
\bibliographystyle{abbrv}
\small
\bibliography{Literaturverzeichnis}

\begin{thebibliography}{10}

\bibitem{ir_almeida_measurement2006}
M.~Almeida, K.~Torrance, and A.~Datta.
\newblock Measurement of optical properties of foods in near- and mid-infrared
  radiation.
\newblock {\em International Journal of Food Properties}, 9(4):651--664, 2006.

\bibitem{rom_twinbuilder_2020}
{ANSYS Inc.}
\newblock {\em ANSYS Twin Builder 2020R2}, 2020.

\bibitem{pm_bird_transport}
R.~B. Bird, W.~E. Stewart, and E.~N. Lightfoot.
\newblock {\em Transport Phenomena}.
\newblock Wiley, 2nd edition, 2006.

\bibitem{pm_datta_porous2007a}
A.~Datta.
\newblock Porous media approaches to studying simultaneous heat and mass
  transfer in food processes. i: Problem formulations.
\newblock {\em Journal of Food Engineering}, 80(1):80 -- 95, 2007.

\bibitem{pm_datta_porous2007b}
A.~Datta.
\newblock Porous media approaches to studying simultaneous heat and mass
  transfer in food processes. ii: Property data and representative results.
\newblock {\em Journal of Food Engineering}, 80(1):96 -- 110, 2007.

\bibitem{pm_datta_toward2016}
A.~Datta.
\newblock Toward computer-aided food engineering: Mechanistic frameworks for
  evolution of product, quality and safety during processing.
\newblock {\em Journal of Food Engineering}, 176:9 -- 27, 2016.

\bibitem{mot_greendeal2030}
{European Commision}.
\newblock {European Green Deal 2030}.
\newblock DOI:10.2775/275924, 2019.

\bibitem{mot_fao_food2011}
{Food and Agriculture Organisation of the United Nations}.
\newblock {\em Global food losses and food waste -- Extent, causes and
  prevention.}
\newblock Rome, 2011.

\bibitem{ir_ibarra_cooked2000}
J.~G.~Ibarra, Y.~Tao, A.~J.~Cardarelli, and J.~Shultz.
\newblock Cooked and raw chicken meat: Emissivity in the mid-infrared region.
\newblock {\em Applied Engineering in Agriculture}, 16(2):143--148, 2000.

\bibitem{pm_halder_surface12}
A.~Halder and A.~K. Datta.
\newblock Surface heat and mass transfer coefficients for multiphase porous
  media transport models with rapid evaporation.
\newblock {\em Food and Bioproducts Processing}, 90(3):475 -- 490, 2012.

\bibitem{diss_simconf19}
M.~Kannapinn and M.~Sch{\"a}fer.
\newblock Endeavouring intelligent process self-control by employing digital
  twin methodology: Proof-of-concept study for cooking applications.
\newblock In {\em Proceedings of CADFEM ANSYS Simulation Conference}, 2019.

\bibitem{pm_kumar_microwave}
C.~Kumar, M.~U.~H. Joardder, T.~W. Farrell, and M.~A. Karim.
\newblock Investigation of intermittent microwave convective drying ({IMCD}) of
  food materials by a coupled {3D} electromagnetics and multiphase model.
\newblock {\em Drying Technology}, 36(6):736--750, 2018.

\bibitem{mot_langsrud_cooking2020}
S.~Langsrud, O.~S{\o}rheim, S.~E. Skuland, V.~L. Almli, M.~R. Jensen, M.~S.
  Gr{\o}vlen, {\O}.~Ueland, and T.~M{\o}retr{\o}.
\newblock Cooking chicken at home: Common or recommended approaches to judge
  doneness may not assure sufficient inactivation of pathogens.
\newblock {\em PLOS ONE}, 15(4):1--27, 04 2020.

\bibitem{rom_ljung_system}
L.~Ljung.
\newblock {\em System Identification - Theory for the User}.
\newblock Prentice Hall PTR, New Jersey, 2nd edition, 1999.

\bibitem{rom_nelles_nonlinear2021}
O.~Nelles.
\newblock {\em Nonlinear System Identification - From Classical Approaches to
  Neural Networks, Fuzzy Models, and Gaussian Processes}.
\newblock Springer International Publishing, 2nd edition, 2020.

\bibitem{pm_ni_multiphase}
H.~Ni.
\newblock {\em Multiphase moisture transport in porous media under internal
  heating of microwaves}.
\newblock PhD thesis, Cornell University, Ithaca, New York, 1997.

\bibitem{mot_zeit2018}
N.~Oberhuber.
\newblock Fachkr{\"a}ftemangel - {Als w{\"a}ren das niedere Arbeiten}, ZEIT
  online, 2018.

\bibitem{mot_ipcc_climate2019_short}
{P. Shukla}.
\newblock {IPCC 2019}: Summary for policymakers.
\newblock In {\em Climate Change and Land: an IPCC special report.}

\bibitem{cm_rabeler_kin2018}
F.~Rabeler and A.~H. Feyissa.
\newblock Kinetic modeling of texture and color changes during thermal
  treatment of chicken breast meat.
\newblock {\em Food and Bioprocess Technology}, 11(8):1495--1504, 2018.

\bibitem{dt_tao_nature}
F.~Tao and Q.~Qi.
\newblock Make more digital twins.
\newblock {\em Nature}, 573:490--491, 2019.

\bibitem{mot_unitednationsgoal}
{United Nations Sustainable Development Group}.
\newblock {sdgs.un.org/goals}, 2021.

\bibitem{dt_verboven_digital2020}
P.~Verboven, T.~Defraeye, A.~K. Datta, and B.~Nicolai.
\newblock Digital twins of food process operations: the next step for food
  process models?
\newblock {\em Current Opinion in Food Science}, 35:79--87, 2020.

\end{thebibliography}

\end{document}